\documentclass[aps,pre,twocolumn,superscriptaddress,showpacs]{revtex4}
\usepackage{amssymb}
\usepackage{epsfig}
\usepackage{amsmath}
\usepackage{times}

\begin{document}

\title{Protecting infrastructure networks from cost-based attacks}
\author{Xingang Wang}
\email[Corresponding author. Email address: ]{wangxg@zju.edu.cn}
\affiliation{Institute for Fusion Theory and Simulation, Zhejiang
University, Hangzhou, China 310027} \affiliation{Temasek
Laboratories, National University of Singapore, 117508, Singapore}
\affiliation{Beijing-Hong Kong-Singapore Joint Centre for Nonlinear
\& Complex Systems (Singapore), National University of Singapore,
Kent Ridge, 119260, Singapore}
\author{Shuguang Guan}
\affiliation{Temasek Laboratories, National University of
Singapore, 117508, Singapore} \affiliation{Beijing-Hong
Kong-Singapore Joint Centre for Nonlinear \& Complex Systems
(Singapore), National University of Singapore, Kent Ridge, 119260,
Singapore}
\author{Choy Heng Lai}
\affiliation{Department of Physics, National University of
Singapore, Singapore 117542} \affiliation{Beijing-Hong
Kong-Singapore Joint Centre for Nonlinear \& Complex Systems
(Singapore), National University of Singapore, Kent Ridge, 119260,
Singapore}

\begin{abstract}
It has been known that heterogeneous networks are vulnerable to the
intentional removal of a small fraction of highly connected or
loaded nodes, which implies that, to protect a network effectively,
a few important nodes should be allocated with more defense
resources than the others. However, if too many resources are
allocated to the few important nodes, the numerous less-important
nodes will be less protected, which, when attacked all together,
still capable of causing a devastating damage. A natural question
therefore is how to efficiently distribute the limited defense
resources among the network nodes such that the network damage is
minimized whatever attack strategy the attacker may take. In this
paper, taking into account the factor of attack cost, we will
revisit the problem of network security and search for efficient
network defense against the cost-based attacks. The study shows
that, for a general complex network, there will exist an optimal
distribution of the defense resources, with which the network is
well protected from cost-based attacks. Furthermore, it is found
that the configuration of the optimal defense is dependent on the
network parameters. Specifically, network that has a larger size,
sparser connection and more heterogeneous structure will be more
benefited from the defense optimization.
\end{abstract}

\date{\today }
 \maketitle

\textbf{Introduction.} -- Modern human societies very much depend on
the efficient functioning and stable operation of complex
infrastructure networks \cite{REV}. Typical examples are electrical
power grids, telecommunication networks, the Internet, and many
transportation systems such as road, railway, and airline networks.
A significant and common feature of these networks is that they all
possess the heterogeneous degree distribution, i.e. they are
scale-free networks (SFN) \cite{BA:1999}. While the adoption of SFN
structure could improve the network performance significantly, e.g.
a shorter average network diameter, it also cause some problems to
the network security. For instance, it has been shown that the
connectivity of a SFN could be largely damaged if a small fraction
of the large-degree nodes are intentionally removed; in contrast, if
the removal is made to the small-degree nodes, the network damage
will be very limited \cite{SFN:TOP}. The robust-yet-fragile property
of SFN is more evident when the intrinsic dynamics of the network
flow is taken into account \cite{SFN:DYN}. This has been shown by a
model of cascade network in Ref. \cite{ML:2002}, where it is found
that, due to the existence of the flow dynamics, the removal of even
a single node could trigger such a large-scale avalanche that only a
small portion of the nodes survive from the cascading failures.
Since practical networks typically carry flows, their securities
against cascading failures thus are of great importance, and have
drawn many attentions in the past years. The topics had been touched
include: Model design \cite{CF:DESN}, damage estimation
\cite{CF:EST}, dynamics characterization \cite{CF:DYNS}, capacity
allocation \cite{CF:ALLT}, topology dependence \cite{CF:CN}, and
cascade control and defense strategies \cite{CF:DEFS}.

\textbf{Problem formulation.} -- While the fragility of SFN to
intentional node removal has been well addressed, so far the studies
have been concentrating on only the case of ``technical" failures,
instead of the real attacks. More specifically, the previous studies
are interested in comparing the extents of the network damage caused
by different implementations of attacks, while neglecting the cost
required in doing so. In a practical situation of network security,
the attacker and the defender are just the two sides of the game.
Their purposes are the same in a sense, i.e., to maximize the gains
with the limited resources. The defender, knowing the important
roles of the large-degree nodes, of course will allocate more
defense resources to them; and the attacker, while desiring to
attack the large-degree nodes, has to scruple about the higher cost
in doing so. Thus in a real attack, the attacker will balance
between the network damage and the attack cost, and search for an
effective attack. For example, by the cost of attacking an important
and well-protected node, the attacker may turn to attacking a number
of non-important and less-protected nodes all together, while the
latter may generate the larger damage. So, before taking an action,
the attacker will do some analysis to the network security, so as to
find the security weak point.

To analyze the network security, the attack usually will design a
series of virtual attacks based on some of the network information,
e.g. the network structure and the defense configuration, and then
evaluate the possible damages caused by the attacks. After a
comparison of the damages, the attack will figure out the most
damaging attack and put it into action. Generally, the virtual
attacks are designed according to two strategies: (1) Concentrating
all the effort to attack a few important and well-protected nodes;
and (2) distributing the effort to a number of non-important and
less-protected nodes. We call the former concentrated attack (CA)
strategy, and the latter distributed attack (DA) strategy. It is
straightforward to see that, if the nodes are equally protected, the
network will be vulnerable to CA; in contrast, if too many defense
resources are allocated to the important nodes, the network will be
vulnerable to DA. Now a challenge faced by the defender is:
\emph{How to optimize the network defense so that the network damage
is minimized whatever attack strategy the attacker may take?}

The problem of cost-based attacks can be formulated as follows. Let
$P=\{p_{i}, i=1,...,N\}$ be the existing defense of an
infrastructure network consisting of $N$ nodes. The defense
resources allocated to node $i$ is $p_{i}$. So the total amount of
the network defense is $R=\sum_{i=1}^{N} p_{i}$. In the current
study, we assume that the attacker has the full knowledge of the
network, including the network topology, the flow dynamics, and the
defense distribution (the general case will be discussed later).
Based on these information, the attacker will scheme out a series of
virtual attacks, $A_n=\{ a_{n,j}, j=1,...,N'\}$, based on either the
CA or DA strategy. In the attack $A_n$, $N'$ out of $N$ nodes in the
network will be selected as the targets, and the cost for removing
target $j$ is denoted by $a_{n,j}$. The total attack cost of $A_n$
therefore is $E_n=\sum_{j=1}^{N'} a_{n,j}=E$, which is identical for
all the attacks pointing to the defense $P$. In general, we have
$E\ll R$. The network damage caused by $A_n$ is denoted by $D_n =
\{b_{n,l}, l=1,...,M\}$, where $\{l\}$ is the set of the failed
nodes due to the attack $A_n$, and $b_{n,l}$ is the amount of
network damage due to the failure of $l$. Then the total network
damage caused by $A_n$ can be quantified: $B_n=\sum_{l=1}^{M}
b_{n,l}$. Evaluating the damage of each of the virtual attacks,
finally the attacker will identify the most devastating attack.

The optimal defense is defined as follows. If the defense resources
are distributed in such a way that all the virtual attacks generate
the same amount of network damage, then this distribution of defense
resources is called the optimal defense, and the network is regarded
as secure to cost-based attacks. Otherwise, if there is difference
between the network damages, the distribution will be considered as
not optimal and the network is regarded as vulnerable to cost-based
attacks. Putting alternatively, if by changing the attack strategy
the attacker can increase the network damage, the network is
considered as not securely protected.

\textbf{The model.} -- We implement the above idea of network
security by a model of cascade network \cite{ML:2002} (the
generalization to the other models are straightforward
\cite{SFN:TOP}). Let $L_i(0)$ be the transmission load (betweenness
centrality) of node $i$, which accounts for the total number of
shortest paths passing though $i$ in the original network \cite{BT}.
Define the node capacity as $C_i=(1+\alpha) L_i(0)$, which
stipulates the maximum load that node $i$ can handle. $\alpha>0$ is
the tolerance parameter. Once a node is attacked, it will be removed
out from the network, together with the links that associate to it.
Because of node removal, the shortest pathes of the network will be
redistributed and, consequently, the load of the remaining nodes
will be updated. In this process, any node which is overloaded, i.e.
$L_i(t)>C_i$, will be removed out from the network. The new removal
will cause a new distribution of the shortest pathes, thus
generating another wave of node failures, and so on and so forth,
till no node is overloaded in the remaining network. To fit this
model into our problem of cost-based attacks, it is necessary to
make a few assumptions. Firstly, it is assumed that the defense
resources have the following power-law distribution,
\begin{equation}
p_i=R \times C_i^{\beta}/C(\beta), \label{DIST}
\end{equation}
where $R$ is the total defense of the network, and
$C(\beta)=\sum_i{C_i^{\beta}}$ is a normalizing factor which is
dependent of the parameter $\beta$. Without loss of generality, here
we set $R=C(\beta=1)$, i.e. the network defense equals the network
capacity. Secondly, it is assumed that the cost for removing a node
is equivalent to the node defense, i.e., $a_i = p_i$. Finally, it is
assumed that the network damage relies on only the removed nodes. In
the current study, the network damage is measured by two quantities:
(1) The size of the largest component in the remaining network, $G$,
and (2) the total capacity of the removed nodes, $B=\sum_{l=1}^{M}
b_{l}$. It is emphasized that these assumptions are made for only
the purpose of illustration. In real applications, they should be
redefined accordingly to the real problems. The key parameter in
this model therefore is $\beta$, which gives the distribution of the
defense resources. When $\beta \ll 0$, the important (high-load)
nodes will be not allocated with the sufficient resources, making
the network vulnerable to CA. In contrast, if $\beta \gg 0$, the
important nodes will be overprotected, making the network vulnerable
to DA. So, to protect the network from cost-based attacks
efficiently, the value of $\beta$ should be properly set.

We next describe the method used in our analysis of the network
security. Noticing the fact that the virtual attacks are divided
into two classes, CA and DA, the network security thus can be
evaluated by considering the two representative attacks. For CA, we
will choose to attack the single node of the largest capacity
(highest protection) in the network; while for DA, \emph{with the
same amount of attack cost}, we will choose to attack a group of
nodes of the smallest capacity (lowest protection). Specifically, if
nodes are ranked by an ascending order of the node capacity, i.e.
$C_1<C_2<\ldots<C_N$, then in CA only node $N$ is attacked, while in
DA nodes from $1$ to $N'$ will be attacked all together. Here $N'$
is a number to be determined by the relation $\sum_{i=1}^{N'}a_i\leq
a_N$. Please note that in a real situation it is possible that the
most devastating attack is neither of the above representative
attacks. However, such a devastating attack, if exists, will be very
dependent on the network particulars, and should be always treated
case by case \cite{JOINTATTK}.

\textbf{Numerical results.} -- To simulate the cost-based attacks,
firstly we generate a SFN by the model proposed in Ref.
\cite{BA:1999}. The network consists of $N=3000$ nodes and has
average degree $\langle k \rangle =4$. The degree distribution
follows a power-law scaling $P(k) \sim k^{\gamma}$, with $\gamma
=-3$. Secondly, we calculate the transmission load of each node and,
according to the value of $\alpha$, calculate the node capacity. For
illustration, here we set $\alpha =0.3$. Then we can obtain the
total defense of the network $R$, which in our model is set to be
the total network capacity, i.e. $R=\sum_i C_i$. Thirdly, we choose
a value for $\beta$ and, according to Eq. (\ref{DIST}), distribute
the defense resources among the nodes. Fourthly, we analyze the
network security by the above mentioned two representative attacks,
and record their damages $G_{1,2}$ and $B_{1,2}$, with the
subscripts 1 and 2 stand for CA and DA, respectively. Finally, by
scanning $\beta$, we are able to figure out the location of the
optimal defense, i.e., the value of $\beta$ where the two attacks
generate the same network damage.

\begin{figure}[tbp]
\begin{center}
\epsfig{figure=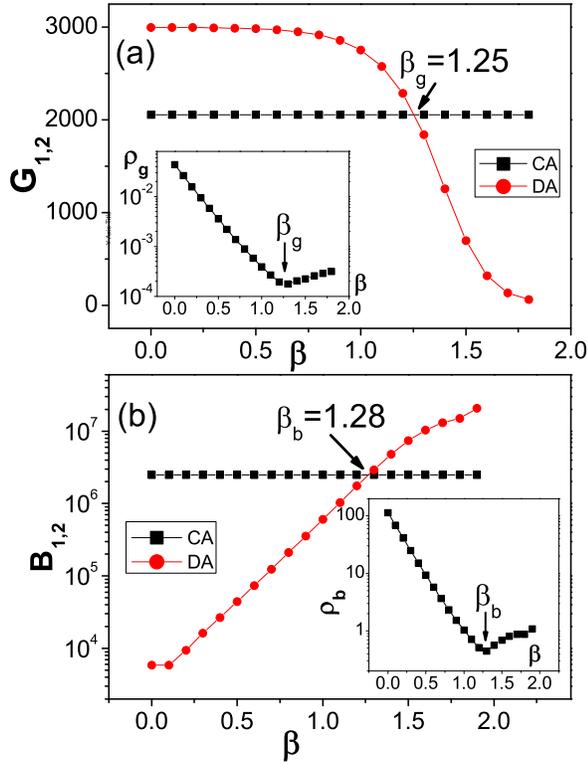,width=0.9\linewidth} \caption{(Color
online) For SFNs of size $N=3000$, average degree $\langle k \rangle
=4$, and tolerance parameter $\alpha =0.3$, the dependence of the
network damage on parameter $\beta$ for CA and DA. (a) $G_{1,2}$
versus $\beta$. The optimal defense is found at about $\beta_g
\approx 1.25$. Inset: $\rho_g$ versus $\beta$. (b) $B_{1,2}$ versus
$\beta$. The optimal defense is found at about $\beta_b \approx
1.28$. Please note the semi-logarithmic plot of $B_{1,2}$. Inset:
$\rho_b$ versus $\beta$. Each date is averaged over 50 network
realizations.} \label{fig:THEORYMODEL}
\end{center}
\end{figure}

The variations of $G$ and $B$ as a function of $\beta$ are plotted
in Figs. 1. For the measurement $G$, the optimal defense is found at
about $\beta_{g} \approx 1.25$ [Fig. 1(a)]; while for the
measurement $B$, the optimal defense is found at about $\beta_{b}
\approx 1.28$ [Fig. 1(b)]. Please note that the optimal defense is
only meaningful to the defender, as it tells how to configure the
defense resources against the cost-based attacks. While for the
attacker, by knowing the specific network defense (the value of
$\beta$), the only task is to figure out which attack is more
damaging, DA or CA. For instance, if the attacker is interested in a
larger damage of network capacity and have learned that the network
defense parameter is $\beta=0.5$, after a comparison of the virtual
attacks, the attacker will find that using CA will cause a larger
damage than DA [Fig. 1(b)].

It is important to note that, in our design of numerical
simulations, CA is always implemented by removing the single node of
the largest capacity. That is the reason why the network damage
caused by CA is constant in Fig. 1. However, as $\beta$ increases,
the cost for removing the largest-capacity node is monotonically
increased, i.e., $E=a_N \sim C_N^{\beta}$. This arises the problem
of attack efficiency, which is defined as the amount of network
damage per unit of the attack cost. For measurement $G$, it is
defined as $\rho_g = (N-G_{M})/E$, with $G_M=\min(G_1,G_2)$; for
measurement $B$, it is defined as $\rho_b = B_{M}/E$, with
$B_M=\max(B_1,B_2)$. Interestingly, it is found that, at the optimal
defense, the attack efficiency is also minimized (the insets of Fig.
1). Now we see that, with the optimal defense, the network is
protected from not only the attack strategy, but also the attack
efficiency.

Physically, the meaning of the optimal defense can be understood as
follows. When $\beta$ is small, say for example $\beta \approx 0$,
the network nodes are equally protected regardless of their
importance level. To generate a large damage, the attacker will
certainly choose to attack the important nodes, i.e. adopting CA. As
$\beta$ increases, more defense resources will be shifted to the
important nodes and, correspondingly, the defense of the
non-important nodes will be weakened. However, as long as $\beta <
\beta_{g,b}$, the damage caused by CA will be still larger to that
of DA. So in this range CA will be always the choice for the
attacker. Nevertheless, as $\beta$ increases, the damage difference
between CA and DA will be gradually narrowed. Then, at the optimal
defense $\beta_{g,b}$, both attacks will generate the same amount of
network damage. Since at this point the attacker can not benefit
from changing between the attacks, the cost-based attacks are
considered as failed. After that, as $\beta$ increases from
$\beta_{g,b}$, the minority important nodes will be overprotected,
and the majority non-important nodes will be less protected. Noticed
of this, the attacker will switch the attack from CA to DA, so as to
achieve a larger damage. In the extreme situation of $\beta \approx
\infty$, all the defense resources will be allocated to the single
node of the largest capacity, while the other nodes of the network
can be easily attacked all together.

\begin{figure}[tbp]
\begin{center}
\epsfig{figure=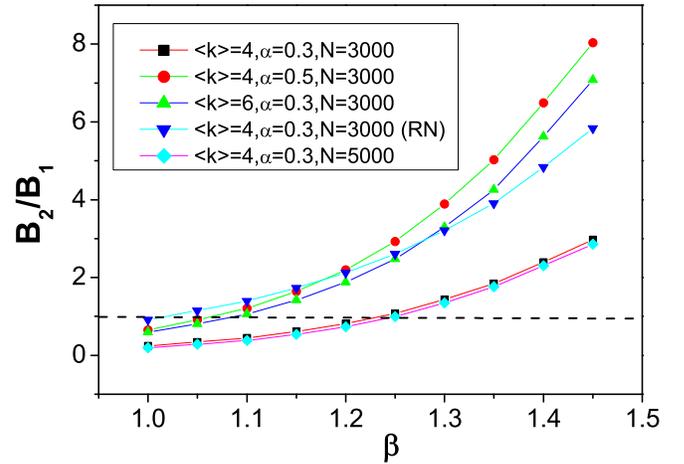,width=\linewidth} \caption{(Color online)
The dependence of $\beta_b$ (characterized by the point where
$B_2(\beta)=B_1(\beta)$) on the network parameters. It is found that
$\beta_b$ is increasing with $N$, but is decreasing with $\alpha$,
$\gamma$, and $\langle k \rangle$. Each data is averaged over 50
network realizations.} \label{fig:PARAMETERDEP}
\end{center}
\end{figure}

As realistic networks have various structures, it is necessary to
check the dependence of the optimal defense to the network
parameters. In particular, we are going to check the dependence of
$\beta_b$ on the following network parameters: The tolerance
parameter $\alpha$, the average degree $\langle k \rangle$, the
degree exponent $\gamma$, and the system size $N$. (The similar
dependence is also valid for $\beta_g$). The numerical results are
plotted in Fig. 2. The general finding is that the value of
$\beta_b$ is increasing with $N$, but is decreasing with $\alpha$,
$\langle k \rangle$, and $\gamma$. (For RN, we have $\gamma
\rightarrow \infty$.) Speaking alternatively, it is the larger,
sparser and heterogeneous networks that will suffer more from the
cost-based attacks and, correspondingly, will be more benefited from
the optimal defense. Since infrastructure networks normally have the
larger size and heterogeneous structure, the studies of optimal
defense thus is of practical concern.

\begin{figure}[tbp]
\begin{center}
\epsfig{figure=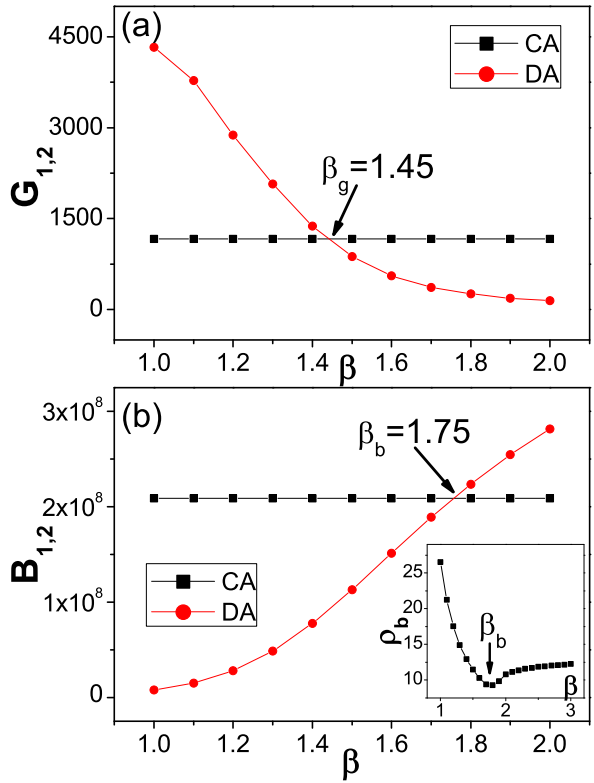,width=0.9\linewidth} \caption{(Color
online) The security analysis for the western U.S. power grid. (a)
The dependence of $G_{1,2}$ on $\beta$, the optimal defense is found
at about $\beta_g \approx 1.45$. (b) The dependence of $B_{1,2}$ on
$\beta$, the optimal defense is found at about $\beta_b \approx
1.75$. Inset: $\rho_b$ versus $\beta$, where $\rho_b$ is minimized
at $\beta_b$. Each data is averaged over 10 attack realizations. For
CA, the top 10 nodes of the highest load are attacked; while for DA,
the nodes are attacked by an ascending order of their capacities.}
\label{fig:POWERGRID}
\end{center}
\end{figure}

\begin{figure}[tbp]
\begin{center}
\epsfig{figure=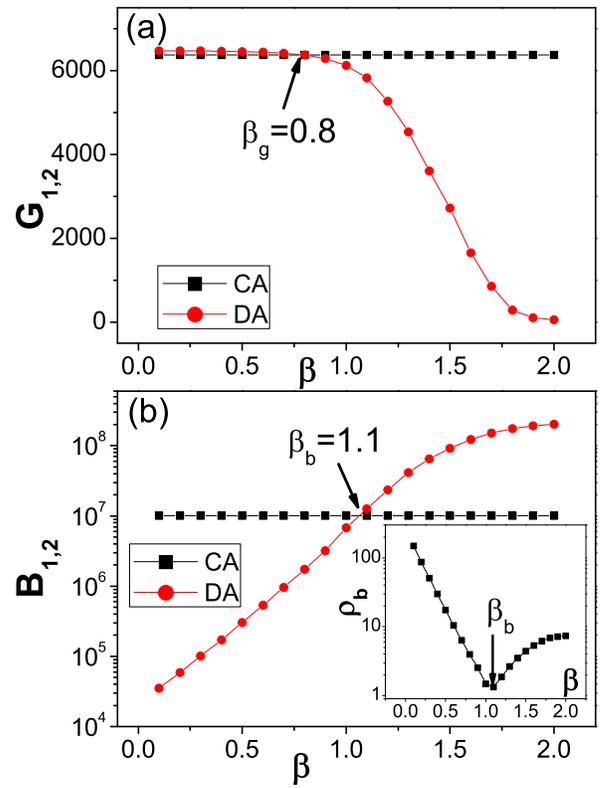,width=0.9\linewidth} \caption{(Color
online) The security analysis for the Internet at the autonomous
level. (a) The dependence of $G_{1,2}$ on $\beta$, $\beta_g \approx
0.8$. (b) The dependence of $B_{1,2}$ on $\beta$, $\beta_b \approx
1.1$. Please note the semi-logarithmic plot of $B_{1,2}$. Inset:
$\rho_b$ versus $\beta$, where $\rho_b$ is minimized at $\beta_b$.
Each data is averaged over 10 attack realizations, just as we did in
Fig. 3.} \label{fig:INTERNET}
\end{center}
\end{figure}

How about the defense of realistic networks? To address this
question, we have analyzed the securities of two typical
infrastructure networks in our society: (1) The electrical power
grid of the western United States \cite{USPOWER}; and (2) the
Internet at the autonomous level \cite{INTERNET}. The power-grid
network consists of $N=4941$ nodes and has average degree $\langle k
\rangle \approx 2.67$, which has been widely used in literature as
an example of cascade network \cite{ML:2002}. The variations of
$G_{1,2}$ as a function of $\beta$ is plotted in Fig. 3(a), where
the optimal defense is found at about $\beta_g \approx 1.45$. In
Fig. 3(b) we plot the dependence of $B_{1,2}$ on $\beta$, where the
optimal defense is found at about $\beta_b \approx 1.75$. Like we
did in Fig. 1, we have also calculated the dependence of the attack
efficiency, $\rho_b$, on the defense parameter $\beta$, where
$\rho_b$ is found to be minimized at $\beta_b$. The Internet we have
employed consists of $N=6474$ nodes and has average degree $\langle
k \rangle \approx 3.88$. The variations of $G_{1,2}$ and $B_{1,2}$
as a function of $\beta$ are plotted in Fig. 4(a) and (b),
respectively. For measurement $G$, the optimal defense is found at
about $\beta_g \approx 0.8$; while for measurement $B$, the optimal
defense is found at about $\beta_b \approx 1.1$. Still, $\rho_b$ is
minimized at $\beta_b$. It is interesting to see that, comparing to
the standard SFN model [Fig. 1] and the power-grid network [Fig. 3],
the Internet is less vulnerable to CA when $\beta < \beta_g$ in
terms of measurement $G$ [Fig. 4(a)]. We attribute this strange
behavior to the unique topology of the Internet, e.g., the modular
structure, the degree correlation, and the hierarchical property.
This also verifies our previous finding of the dependence of optimal
defense on network parameters [Fig. 2].

\textbf{Discussion and conclusion.} -- The main purpose of the
present study is to highlight the \emph{variability and flexibility}
of the network attacks in the real situation, so as to bring a
caution to the defense of complex networks. Our main finding is
that, if the defense resources of a network are not well
distributed, the attacker could be benefited from choosing between
the attack strategies. In showing this, we had employed the simple
model of cascade network and made a few assumptions on the network
defense and attack, which, when used to model the real situations,
should be (carefully) modified and redefined. For instance, it has
been shown recently that, as a balance of network robustness and
frangibility, the relationship between node capacity and load could
be nonlinear \cite{KM:2008}. This indicates that, to analyze the
security of such a network, the constant tolerance parameter used in
the current model should be modified. This kind of modifications,
however, will not change the general picture of optimal defense. The
fact is that, as long as the cost factor of network attack is
counted, optimal defense will exist and be an important issue in
network security.

A point which should be specially addressed is that the current
model requires a full knowledge of the network, including the detail
information about the network structure and flow dynamics. These
information, while is available for some public systems such as the
power-grid \cite{POWERGRID} and the Internet, is difficult to obtain
for the secret networks, say, for example, the terror and Mafia
networks. In a secret network, the important nodes, which possess
the larger degree and have higher ranks in the hierarchy, are
usually well covered and difficult to identify. This arouses the
problem of attacking probability, a question investigated by Gallos
\emph{et al.} very recently \cite{ATTACKPROB}. In that study, the
probability of removing a node is determined by three factors: The
node degree $k$, the intrinsic network vulnerability $\alpha'$, and
the node knowledge $\alpha{''}$. There a key finding is that, as the
information of the important nodes be gradually exposed (increasing
the value of $\alpha{''}$), the fraction of nodes needed to break
the network will be quickly decreased. Here, an interesting thing is
that, if we regard the cover of the network information as an
approach of network defense, the study of Ref. \cite{ATTACKPROB} and
the present work have essentially the same basis. In particular, if
we replace the parameter $\beta$ in Eq. (\ref{DIST}) by a new
parameter $(\alpha'+\alpha^{''}) / \kappa$ ($\kappa \approx 1.6$ is
the exponent that characterizes the relationship between the node
capacity and degree \cite{SFN:BETDIS}), then the node defense
defined in Eq. (\ref{DIST}) is just the reciprocal of the node
vulnerability defined in Ref. \cite{ATTACKPROB}. For this reason, we
may say that the study in Ref. \cite{ATTACKPROB} is a special case
of the cost-based attacks proposed in the present work. Despite of
this point of similarity, the two studies are actually dealing with
very different problems. Simply speaking, the study of Ref.
\cite{ATTACKPROB} is focusing on the scale of network damage, in
which the attack cost (information discovery) is variable and the
attack strategy is always fixed to CA; in contrast, the current
study is dealing with the situation of variable attack strategy and
fixed attack cost, i.e., it is a question about network optimization
\cite{MT:2007}.

Summarizing up, we have proposed the idea of cost-based attacks on
complex networks and investigated the problem of optimal network
defense. Different from previous studies, here we emphasize the
initiative and flexibility of the attacker in implementing the
attacks, which is a solid step forward to the realistic situations.
We hope this study could stimulate new thinking to the security of
complex networks, and give indications to the design and defense of
infrastructure networks.

XGW is supported by the National Natural Science Foundation of China
under Grant No. 10805038.

\end{document}